# Dynamic Symmetries and Entropic Inequalities in the Probability Representation of Quantum Mechanics

Margarita A. Man'ko and Vladimir I. Man'ko

*P. N. Lebedev Physical Institute, Leninskii Prospect 53, Moscow 119991, Russia*

**Abstract.** The probability representation of quantum and classical statistical mechanics is discussed. Symplectic tomography, center-of-mass tomography, and spin tomography are studied. The connection of tomographic probabilities with dynamic symmetries like symplectic group is considered. Entropic uncertainty relations and inequalities for spin tomograms are reviewed.

## INTRODUCTION

The concepts of state in quantum and classical mechanics seem to be different. We show that one can find the formulation of quantum mechanics where the concept of state does not differ from the concept of state in classical statistical mechanics. The approach we develop is based on the group-theoretical methods and the properties of system symmetries, e.g., linear canonical transforms [1, 2, 3] belonging to symplectic group of dynamic symmetry of such systems as harmonic oscillator [4] (see also [5]). The distinguished pioneer contributions to such symmetry application in quantum mechanics were done by Marcos Moshinsky.

The aim of this article is to present the material of lectures given by the authors at the XL Latin American School of Physics *Symmetries in Physics* in memory of Marcos Moshinsky and devoted to the tomographic-probability representation of quantum mechanics [6, 7].

There exist different formulations of quantum mechanics [8], including the Schrödinger approach [9], the Moyal approach [10] based on the notion of Wigner function [11], and the Feynmann path integral approach [12]. In all these formulations, the basic concept of quantum state is not the concept used in classical statistical mechanics where the state in associated with the probability density on the phase space of the system. Quantum analogs of different kinds of such probability densities, in addition to the Wigner function, were suggested in the form of Husimi–Kano $Q$-function [13, 14] or the Glauber–Sudarshan $P$-function [15, 16].

In view of the position–momentum uncertainty relation, a joint probability distribution on the phase space for such observables does not exist. Consequently, one cannot associate the fair probability distribution on the phase space of measurable position and momentum with a quantum state. The tomographic-probability approach for de-scribing quantum states (called the probability representation of quantum mechanics) was suggested for the both continuous variables [17] and discrete variables like spin or qubits [18, 19] (see also the development in [20, 21, 22, 23, 24, 25, 26, 27]).

Other kinds of tomographies for continuous variables were considered in [28]. The discussion of different aspects of tomographic approaches can be found in [29, 30, 31, 32, 33, 34, 35].

The tomographic approach in classical mechanics was introduced in [36, 37].

In [38], it was shown that the quantumness and classicality of the system behavior can be considered in the tomographic probability representation of quantum mechanics. The entropic aspects of tomographic approach were studied in [39, 40, 41, 42, 43, 44, 45, 46, 47, 48].

The tomographic approach can be developed using the dynamic symmetries [5] of quantum systems. The hydrogen atom has the dynamic symmetry which is $O(4,2)$ group [49]. In tomography, the symplectic group $Sp(2n,R)$ plays an important role [50]. In [50], it was suggested to use homodyne measuring of optical tomograms to check the uncertainty relations.

The paper is organized as follows.

First, the tomography of classical states is reviewed, and the symplectic tomography of quantum-particle states is considered. Also the symplectic and center-of-mass tomographies of multimode systems are studied. In the second part of this paper, the spin tomography is presented, and entropies associated with tomographic-probability distributions are discussed. In the third part of this paper, quantumness tests are employed, and finally the conclusions and prospectives are shown.

## TOMOGRAPHIC-PROBABILITY DISTRIBUTION OF CLASSICAL PARTICLE

The states of a particle in classical statistical mechanics are described by the probability distribution function $f(q,p)$ on the phase space, where $q$ and $p$ are, respectively, the classical particle's position and momentum considered in a reference frame with axes $q$ and $p$ (see Figure 1).

Given the probability density $f(q,p)$ on the phase space. The function $f(q,p)$, due to the physical meaning of the probability distribution, is nonnegative and normalized

$$\int f(q,p)\,dq\,dp = 1. \qquad (1)$$

Let us calculate the marginal probability density of the particle's position $X$ in a rotated reference frame on the phase space with new axes $q'$ and $p'$. One has the expression for position $X$ in rotated reference frame as

$$X = q\cos\theta + p\sin\theta, \qquad (2)$$

where $\theta$ is the rotation angle. One can see that for $\theta = 0$, $X = q$ and for $\theta = \pi/2$, $X = p$.

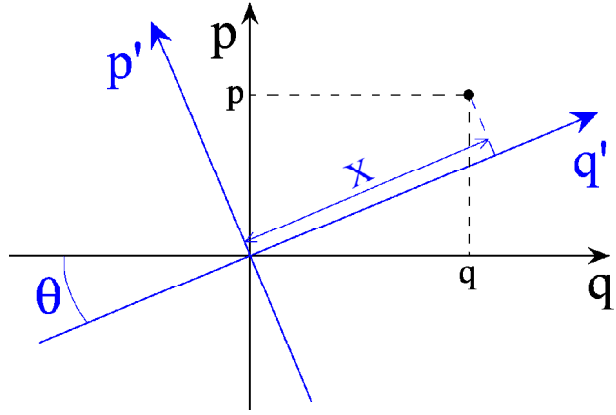

**FIGURE 1.** Particle's position $X$ measured in rotated reference frame in the phase space.

The marginal probability density $w(X, \theta)$ (called optical tomogram in quantum optics, but we will use this name also in classical statistical mechanics) reads

$$w(X,\theta) = \langle \delta(X - q\cos\theta - p\sin\theta) \rangle = \int f(q,p)\delta(X - q\cos\theta - p\sin\theta)\,dq\,dp. \quad (3)$$

Now we introduce another tomogram (related to the optical tomogram) accompanying the rotation of reference frame in the phase space by scaling the position and momentum before the rotation. Namely, we consider the marginal probability density denoted as $M(X,\mu,\nu)$ of the particle's position $X$ in a reference frame on the phase space, which first was rescaled (i.e., $q \to sq$ and $p \to s^{-1}p$, where $s$ is a real scaling parameter) and then was rotated; it reads

$$M(X,\mu,\nu) = \langle \delta(X - \mu q - \nu p) \rangle = \int f(q,p)\delta(X - \mu q - \nu p)\,dq\,dp, \quad (4)$$

where $\mu = s\cos\theta$ and $\nu = s^{-1}\sin\theta$.

The probability distribution $M(X,\mu,\nu)$ is called symplectic tomogram of the classical particle's state. It is normalized $\int M(X,\mu,\nu)\,dX = 1$ due to the property of delta-function $\int \delta(X - \mu q - \nu p)\,dX = 1$ and the normalization of the distribution $f(q,p)$ on the phase space.

Due to homogeneity of the Dirac delta-function, i.e., $\delta(\lambda x) = |\lambda|^{-1}\delta(x)$, the symplectic tomogram is also homogeneous function, $M(\lambda X, \lambda \mu, \lambda \nu) = \frac{1}{|\lambda|} M(X,\mu,\nu)$. Thus, one has the connection between the optical and symplectic tomograms, due to the homogeneity property, i.e.,

$$w(X,\theta) = M(X,\cos\theta,\sin\theta), \quad (5)$$

$$M(X,\mu,\nu) = \frac{1}{\sqrt{\mu^2 + \nu^2}} w\left(\frac{X}{\sqrt{\mu^2 + \nu^2}}, \tan^{-1}\frac{\nu}{\mu}\right). \quad (6)$$

Formulae for optical tomogram, given by its definition (3) and symplectic tomogram, given by its definition (4), turn out to coincide with the well-known integral Radon transform [51, 52] of the function of two variables $f(q,p)$, which has the inverse. The function of two variables can be an arbitrary function. If the physical meaning of the function of two variables $f(q,p)$ is such that it is considered as nonnegative normalized probability distribution on the phase space, the inverse Radon transform provides this distribution given in terms of the tomographic nonnegative probability distribution $M(X,\mu,\nu)$ as follows:

$$f(q,p) = \frac{1}{4\pi^2} \int M(X,\mu,\nu) e^{i(X-\mu q-\nu p)} dX\,d\mu\,d\nu \geq 0. \tag{7}$$

The probability distribution $M(X,\mu,\nu)$ can be used to calculate momenta of the random variables $q$ and $p$. In fact, due to the physical meaning of the marginal probability distribution $M(X,\mu,\nu)$, one has

$$\langle q^n \rangle = \int M(X,1,0) X^n\,dX, \tag{8}$$

$$\langle p^n \rangle = \int M(X,0,1) X^n\,dX. \tag{9}$$

The momenta can be obtained using the characteristic function $\xi(k,\mu,\nu)$ of the distribution $M(X,\mu,\nu)$ as coefficients of the Taylor expansion, given as

$$\xi(k,\mu,\nu) = \langle e^{ikX} \rangle = \int e^{ikX} M(X,\mu,\nu)\,dX. \tag{10}$$

The probability distribution $M(X,\mu,\nu)$ can be expressed in terms of the Fourier transform of the characteristic functions as follows:

$$M(X,\mu,\nu) = \frac{1}{2\pi} \int \xi(k,\mu,\nu) e^{-ikX}\,dk. \tag{11}$$

Given a nonnegative normalized homogeneous function $\widetilde{M}(X,\mu,\nu)$. If the function has nonnegative Fourier integral (7), it belongs to a set of symplectic tomograms determining the states of a classical particle.

## TOMOGRAPHIC-PROBABILITY DISTRIBUTION OF QUANTUM PARTICLE

Now we introduce tomographic probability distributions (tomograms) of a quantum particle. We introduce both quantum tomograms: optical tomogram $w(X,\theta)$ and symplectic tomogram $M(X,\mu,\nu)$. We use the same notation for these tomograms like in the case of classical particle. Below we take Planck's constant $\hbar = 1$.

The quantum particle's state can be described by the tomogram obtained using formula (3) for classical tomogram with averaging the delta-function. So, we start from this formula keeping only the form with averaging but then replacing in this form the numbers $q$ and $p$ by corresponding operators. Thus, the position is replaced by the position operator $q \to \hat{q}$ and the momentum is replaced by the momentum operator $p \to \hat{p}$. Also the classical averaging has to be replaced with averaging by means of the quantum-state density operator $\hat{\rho}$.

For photon states, the photon quadrature components play the role of position $q$ and momentum $p$. Then, for the photon quantum state in quantum optics, the optical tomogram is defined as

$$w(X,\theta) = \langle \delta(X - \hat{q}\cos\theta - \hat{p}\sin\theta)\rangle. \tag{12}$$

The average means that we replaced the probability distribution $f(q,p)$ in the definition of classical optical tomogram by the density operator, i.e., $f(q,p) \to \hat{\rho}$ and applied the formula for average of the operator $\hat{A}$ of the form $\langle \hat{A}\rangle = \mathrm{Tr}(\hat{\rho}\hat{A})$. The definition of optical tomogram (12) can be done in a more known form (see [53, 54]), which uses the Wigner function $W(q,p)$ of the photon quantum state

$$w(X,\theta) = \int W(q,p)\delta(X - q\cos\theta - p\sin\theta)\frac{dq\,dp}{2\pi}. \tag{13}$$

One can see that the optical tomogram of classical particle is given by the same formula (3) with replacement $W(q,p)/2\pi \to f(q,p)$. The above optical tomogram of the photon quantum state is measured by homodyne detector [55].

The symplectic tomogram of quantum state is given by the classical formula (4) with the same replacements $q \to \hat{q}$ and $p \to \hat{p}$, i.e.,

$$M(X,\mu,\nu) = \langle \delta(X - \mu\hat{q} - \nu\hat{p})\rangle. \tag{14}$$

Here again we use the above form and averaging is understood as quantum averaging with density operator $\hat{\rho}$. Due to the property of delta-function, the tomogram is normalized $\int M(X,\mu,\nu)\,dX = 1$ for any parameters $\mu$ and $\nu$.

One can express, in view of the definition of the quantum symplectic tomogram for the pure quantum state $|\psi\rangle\langle\psi|$, the symplectic tomogram in terms of the wave function $\psi(y)$ in the position representation, the symplectic tomogram reads (see [56, 57])

$$M(X,\mu,\nu) = \frac{1}{2\pi|\nu|}\left|\int \psi(y)\exp\left[i\left(\frac{\mu}{2\nu}y^2 - \frac{Xy}{\nu}\right)\right]dy\right|^2. \tag{15}$$

The quantum tomogram $M(X,\mu,\nu)$ determines the density operator $\hat{\rho}$ by the formula analogous to the classical formula (7) for reconstructing the probability distribution $f(q,p)$ on the phase space, but with the replacement $f(q,p) \to \hat{\rho}$, $q \to \hat{q}$, $p \to \hat{p}$, and $1/4\pi^2 \to 1/2\pi$ by the formula obtained in [21], i.e.,

$$\hat{\rho} = \frac{1}{2\pi}\int M(X,\mu,\nu)e^{i(X - \mu\hat{q} - \nu\hat{p})}\,dX\,d\mu\,d\nu. \tag{16}$$

One can see that inverse Radon transform (7) for classical symplectic tomogram $M(X,\mu,\nu)$ coincides with its Fourier transform. Reconstruction formula (16) for quantum density operator $\hat{\rho}$ has the form of "quantized" Fourier transform of quantum symplectic tomogram $M(X,\mu,\nu)$.

The density operator of quantum state must be nonnegative. If nonnegative normalized homogeneous function $\widetilde{M}(X,\mu,\nu)$ has the "quantized" Fourier integral of the form (16), which is nonnegative operator, the function belongs to the set of quantum tomograms related to quantum states.

We summarize the notion of classical or quantum states in terms of tomograms $M(X,\mu,\nu)$ in the tomographic probability representation as follows.

The states in both classical and quantum mechanics are determined by nonnegative normalized homogeneous probability distributions $M(X,\mu,\nu)$ (tomograms) depending on a random variable $X$ and real parameters $\mu$ and $\nu$.

We list again the properties of tomograms:

- Nonnegativity
$$M(X,\mu,\nu) \geq 0; \tag{17}$$

- Normalization
$$\int M(X,\mu,\nu)\,dX = 1; \tag{18}$$

- Homogeneity
$$M(\lambda X, \lambda\mu, \lambda\nu) = \frac{1}{|\lambda|} M(X,\mu,\nu). \tag{19}$$

Property (19) means that

$$\left(X\frac{\partial}{\partial X} + \mu\frac{\partial}{\partial \mu} + \nu\frac{\partial}{\partial \nu} + 1\right) M(X,\mu,\nu) = 0. \tag{20}$$

The difference of classical and quantum tomograms is the following one.

In addition to (17)–(19), the classical-state tomogram must satisfy the condition of nonnegativity of its Fourier transform,

$$\int M(X,\mu,\nu) e^{i(X-\mu q - \nu p)}\,dX\,d\mu\,d\nu \geq 0. \tag{21}$$

In addition to (17)–(19), the quantum-state tomogram must satisfy the condition of nonnegativity of the integral determining the operator

$$\int M(X,\mu,\nu) e^{i(X-\mu \hat{q} - \nu \hat{p})}\,dX\,d\mu\,d\nu \geq 0, \tag{22}$$

where $\hat{q}$ and $\hat{p}$ are the position and momentum operators.

The different properties (21) and (22) take place due to the following reasons.

In classical mechanics, integral (21) is interpreted as the probability density $f(q,p)$ on the phase space which must be nonnegative.

In quantum mechanics, integral (22) is interpreted as the density operator $\hat{\rho}$, which must be nonnegative, i.e., it must have only nonnegative eigenvalues.

Conditions (21) and (22) are different conditions for tomogram $M(X,\mu,\nu)$ in spite that, in the tomographic-probability representation, the states of classical and quantum particles are described by the same object – tomogram $M(X,\mu,\nu)$.

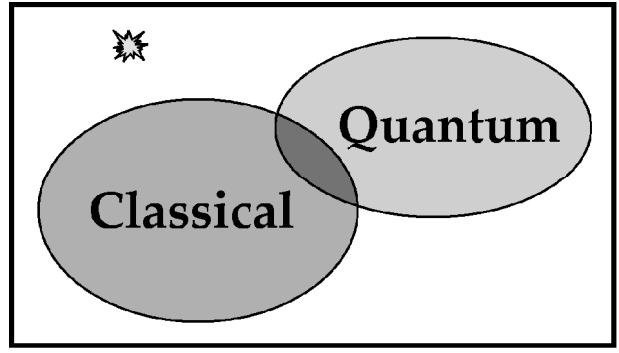

**FIGURE 2.** "Box" containing all nonnegative normalized homogeneous tomograms $M(X,\mu,\nu)$. Intersecting subsets of classical and quantum tomograms and the point outside the subset corresponding neither classical nor quantum tomograms.

The sets of classical tomograms and quantum tomograms are different sets. Both sets have a common part of Gaussian tomograms satisfying simultaneously conditions (21) and (22).

The classical tomograms belong to the set of functions $M(X,\mu,\nu)$ (classical set) obeying the properties of nonnegativity, normalization, homogeneity, and (21).

The quantum tomograms belong to the set of functions $M(X,\mu,\nu)$ (quantum set) obeying the properties of nonnegativity, normalization, homogeneity, and (22).

The set of all tomograms and relation of quantum and classical domains in this set is shown in Figure 2. There are tomograms (see point in "box" outside of the classical and quantum domains) which correspond neither quantum nor classical state.

Optical tomograms $w(X,\theta)$ are given in terms of symplectic tomograms – the functions $M(X,\mu,\nu)$ as $w(X,\theta) = M(X,\cos\theta,\sin\theta)$.

For classical states, the optical tomogram is determined by this formula with the function $M(X,\mu,\nu)$ belonging to the classical set of symplectic tomograms.

For quantum states, the optical tomogram is determined by this formula with the function $M(X,\mu,\nu)$ belonging to the quantum set of symplectic tomograms.

The quantum optical and symplectic tomograms satisfy the same formulae (5) and (6) like the classical tomograms. This means that measuring quantum optical tomograms $w(X,\theta)$ by homodyne detector implies measuring the symplectic tomogram.

The classical evolution equations move the tomogram in the box only inside the classical domain. The quantum evolution equations move the tomogram only inside the quantum domain. The intersection of the classical and quantum domains is invariant for the both classical and quantum evolution equations with quadratic in position and momentum Hamiltonians.

For an arbitrary operator $\hat{A}$, one can introduce the tomographic symbol of the operator extending formula (4)

$$M_A(X,\mu,\nu) = \text{Tr}\,\hat{A}\,\delta(X - \mu\hat{q} - \nu\hat{p}). \tag{23}$$

The symbol $w_A(X,\mu,\nu)$ determines the operator $\hat{A}$ in view of the inverse quantized Fourier transform (6)

$$\hat{A} = \frac{1}{2\pi} \int M_A(X,\mu,\nu) e^{i(X-\mu\hat{q}-\nu\hat{p})} dX\, d\mu\, d\nu. \tag{24}$$

It is worth noting that in [17] the tomogram of photon quantum state $w(X,\mu,\nu,a)$ was introduced as follows:

$$w(X,\mu,\nu,a) = \mathrm{Tr}\,\hat{\rho}\,\delta(X-\mu\hat{q}-\nu\hat{p}-a), \tag{25}$$

where $a$ is a shift parameter. The tomogram is the probability of shifted squeezed and rotated quadrature $X$ – the eigenvalue of the operator

$$\hat{X} = \hat{S}\hat{R}\hat{T}\hat{q}\hat{T}^\dagger\hat{R}^\dagger\hat{S}^\dagger. \tag{26}$$

The dynamic symmetry used in this construction is the inhomogeneous symplectic group $ISp(2,R)$, and the generators of the Lie algebra of this group contain the momentum operator, i.e., the shift operator is

$$\hat{T} = \exp(i\hat{p}a), \tag{27}$$

as well as the rotation operator

$$\hat{R} = \exp\left[i\theta\left(\frac{\hat{p}^2}{2} + \frac{\hat{q}^2}{2}\right)\right] \tag{28}$$

and squeezing operator

$$\hat{S} = \exp\left[i\frac{\ln s}{2}(\hat{q}\hat{p} + \hat{p}\hat{q})\right]. \tag{29}$$

The action of these three operators on the position operator $\hat{q}$ in (26) yields

$$\hat{X} = s\hat{q}\cos\theta + s^{-1}\hat{p}\sin\theta + a. \tag{30}$$

The density operator reconstructed by inverting formula (25) reads

$$\hat{\rho} = \frac{1}{2\pi}\int w(X,\mu,\nu,a) e^{i(X-\mu\hat{q}-\nu\hat{p}-a)} \delta(a)\, dX\, d\mu\, d\nu\, da. \tag{31}$$

The tomogram satisfies the homogeneity condition which in differential form reads

$$\left(X\frac{\partial}{\partial X} + \mu\frac{\partial}{\partial \mu} + \nu\frac{\partial}{\partial \nu} + a\frac{\partial}{\partial a} + 1\right) w(X,\mu,\nu,a) = 0. \tag{32}$$

The shift parameter $a$ provides redundant information for reconstructing the density operator, since only the dependence on the local oscillator phase of the tomogram is sufficient for reconstructing the Wigner function of the state [17], but all three parameters $\theta$, $s$, and $a$ of the symplectic transform (30) reflect the properties of dynamic symmetry of the system's phase space. The squeeze tomography, providing the dependence of tomogram only on the squeezing parameter was introduced in [22], in view of injective map of the density operator onto squeeze tomogram.

The trace of any operator $\hat{A}$ can be calculated in terms of its symbol

$$\mathrm{Tr}\hat{A} = \int M_A(X,\mu,\nu)e^{iX}\delta(\mu)\delta(\nu)\,dX\,d\mu\,d\nu. \tag{33}$$

For symplectic tomographic symbols, this formula is compatible with the following formula:

$$\mathrm{Tr}\hat{A} = \int M_A(X,\mu,\nu)\,dX. \tag{34}$$

The tomogram of the vacuum state $\hat{\rho} = |0\rangle\langle 0|$ reads

$$M_0(X,\mu,\nu) = \frac{1}{\sqrt{\pi(\mu^2+\nu^2)}}\exp\left(-\frac{X^2}{\mu^2+\nu^2}\right). \tag{35}$$

Both formulae (33) and (34) provide $\mathrm{Tr}|0\rangle\langle 0| = 1$. The transition amplitude $\mathrm{Tr}|\psi_1\rangle\langle \psi_2|$ can be expressed in terms of the tomographic symbol of the operator $|\psi_1\rangle\langle \psi_2|$, which is

$$M_{12}(X,\mu,\nu) = \frac{1}{2\pi|\nu|}\int \psi_1(y)\psi_2^*(z)\exp\left[\frac{i\mu}{2\nu}(y^2-z^2)-\frac{iX}{\nu}(y-z)\right]dy\,dz. \tag{36}$$

One can see that

$$\int M_{12}(X,\mu,\nu)\,dX = \int \psi_1(y)\psi_2^*(y)\,dy = \langle \psi_2|\psi_1\rangle. \tag{37}$$

The same result can be obtained using expression (33).

The formula for $\mathrm{Tr}(\hat{A}\hat{B})$ can be given in terms of symbols of the operators $\hat{A}$ and $\hat{B}$ as follows:

$$\mathrm{Tr}(\hat{A}\hat{B}) = \frac{1}{2\pi}\int M_A(X,\mu,\nu)M_B(Y,\mu,\nu)e^{i(X-Y)}dX\,dY\,d\mu\,d\nu. \tag{38}$$

For two quantum states with density operators $\hat{\rho}_1$ and $\hat{\rho}_3$, one has the fidelity

$$\mathrm{Tr}(\hat{\rho}_1\hat{\rho}_2) = \frac{1}{2\pi}\int M_1(X,\mu,\nu)M_2(Y,\mu,\nu)\cos(X-Y)\,dX\,dY\,d\mu\,d\nu. \tag{39}$$

For arbitrary state tomograms, one has the property

$$\int M_1(X,\mu,\nu)M_2(Y,\mu,\nu)\sin(X-Y)\,dX\,dY\,d\mu\,d\nu = 0. \tag{40}$$

In optical experiments, one can measure optical tomograms $w_1(x,\theta)$ and $w_2(y,\theta)$ for two different states $\hat{\rho}_1$ and $\hat{\rho}_2$. We obtain the condition for optical tomograms, where $x$ and $y$ are homodyne quadratures and $\theta$ the local oscillator phase.

Let us introduce the joint probability distribution of two quadratures averaged over the local oscillator phase

$$P_{12}(x,y) = \frac{1}{2\pi} \int_0^{2\pi} w_1(x,\theta) w_2(y,\theta) \, d\theta. \tag{41}$$

The probability distribution is nonnegative and normalized

$$\int P_{12}(x,y) \, dx \, dy = 1. \tag{42}$$

The probability distribution provides the marginal distribution $\mathscr{P}(b)$ in rotated reference frame of two quadratures $x$ and $y$, namely, the frame with coordinates $a = (x+y)/\sqrt{2}$ and $b = (x-y)/\sqrt{2}$. The marginal probability distribution $\mathscr{P}(b)$ reads

$$\mathscr{P}(b) = \int P_{12}\left(\frac{a+b}{\sqrt{2}}, \frac{a-b}{\sqrt{2}}\right) da. \tag{43}$$

The characteristic function for the probability distribution $\mathscr{P}(b)$ is

$$\xi(\lambda) = \langle e^{i\lambda b} \rangle = \int_{-\infty}^{\infty} e^{i\lambda b} \mathscr{P}(b) \, db. \tag{44}$$

It determines momenta of the random variable $b$.

The fidelity $F_{12} = \mathrm{Tr}\,\hat{\rho}_1 \hat{\rho}_2$ is expressed is terms of the characteristic function as follows:

$$F_{12} = \frac{1}{2} \int_0^{\infty} \lambda \xi(\lambda) \, d\lambda. \tag{45}$$

It must satisfy the inequalities

$$0 \leq \frac{1}{2} \int_0^{\infty} \lambda \xi(\lambda) \, d\lambda \leq 1. \tag{46}$$

Experimental optical tomogram must fulfill the condition

$$\mathrm{Im} \int_0^{\infty} \lambda \xi(\lambda) \, d\lambda = 0. \tag{47}$$

This means that

$$\int_0^{2\pi} d\theta \int_0^{\infty} dr \iint_{-\infty}^{\infty} dx \, dy \, w_1(x,\theta) w_2(y,\theta) r \sin[r(x-y)] = 0, \tag{48}$$

where $w_1(x,\theta)$ and $w_2(y,\theta)$ are experimental optical tomograms of two states with density matrices $\hat{\rho}_1$ and $\hat{\rho}_2$.

The fidelity can be expressed also in the form

$$F_{12} = \frac{1}{2\pi} \int_0^{2\pi} d\theta \int_0^{\infty} dr \iint_{-\infty}^{\infty} dx \, dy \, w_1(x,\theta) w_2(y,\theta) r \cos[r(x-y)]. \tag{49}$$

For pure state, $\hat{\rho}_1 = \hat{\rho}_2$ and $\hat{\rho}_1^2 = \hat{\rho}_1$, integral (49) must be equal 1.

# TOMOGRAPHIES OF MULTIMODE PHOTON STATES

There exist different possibilities to introduce tomographic-probability distributions for multimode ($N$-mode) photon states with density operator $\hat{\rho}$. The symplectic tomogram of classical state $M(\vec{X}, \vec{\mu}, \vec{\nu})$ is determined by the formula

$$M(\vec{X},\vec{\mu},\vec{\nu}) = \langle \prod_{k=1}^{N} \delta(X_k - \mu_k q_k - \nu_k p_k) \rangle = \int f(\vec{q},\vec{p}) \prod_{k=1}^{N} \delta(X_k - \mu_k q_k - \nu_k p_k) \, d\vec{q}\, d\vec{p}, \tag{50}$$

where $\vec{X} = (X_1, X_2, \ldots, X_N)$, $\vec{\mu} = (\mu_1, \mu_2, \ldots, \mu_N)$, and $\vec{\nu} = (\nu_1, \nu_2, \ldots, \nu_N)$.

The tomogram of quantum state with the density operator $\hat{\rho}$ is defined analogously,

$$M(\vec{X},\vec{\mu},\vec{\nu}) = \langle \prod_{k=1}^{N} \delta(X_k - \mu_k \hat{q}_k - \nu_k \hat{p}_k) \rangle = \mathrm{Tr}\,\hat{\rho} \prod_{k=1}^{N} \delta(X_k - \mu_k \hat{q}_k - \nu_k \hat{p}_k). \tag{51}$$

The tomograms obey to the equations

$$\left( X_k \frac{\partial}{\partial X_k} + \mu_k \frac{\partial}{\partial \mu_k} + \nu_k \frac{\partial}{\partial \nu_k} + 1 \right) M = 0. \tag{52}$$

One can also introduce the center-of-mass tomogram of classical state

$$M_{\mathrm{cm}}(X,\vec{\mu},\vec{\nu}) = \langle \delta(X - \vec{\mu}\vec{q} - \vec{\nu}\vec{p}) \rangle = \int f(\vec{q},\vec{p})\delta(X - \vec{\mu}\vec{q} - \vec{\nu}\vec{p})\, d\vec{q}\, d\vec{p} \tag{53}$$

and the center-of-mass tomogram of quantum state

$$M_{\mathrm{cm}}(X,\vec{\mu},\vec{\nu}) = \langle \delta\left(X - \vec{\mu}\hat{\vec{q}} - \vec{\nu}\hat{\vec{p}}\right) \rangle = \mathrm{Tr}\,\hat{\rho}\, \delta\left(X - \vec{\mu}\hat{\vec{q}} - \vec{\nu}\hat{\vec{p}}\right). \tag{54}$$

There exist inverse maps for formulae (50)–(54).

For classical symplectic tomogram, one has

$$f(\vec{q},\vec{p}) = \frac{1}{(2\pi)^{2N}} \int M(\vec{X},\vec{\mu},\vec{\nu}) \left( \prod_{k=1}^{N} e^{i(X_k - \mu_k q_k - \nu_k p_k)} \right) d\vec{X}\, d\vec{\mu}\, d\vec{\nu}. \tag{55}$$

For quantum symplectic tomogram, the density operator can be reconstructed using quantized Fourier transform

$$\hat{\rho} = \frac{1}{(2\pi)^{N}} \int M(\vec{X},\vec{\mu},\vec{\nu}) \left( \prod_{k=1}^{N} e^{i(X_k - \mu_k \hat{q}_k - \nu_k \hat{p}_k)} \right) d\vec{X}\, d\vec{\mu}\, d\vec{\nu}. \tag{56}$$

The center-of-mass tomogram determines the distribution on the phase space for classical system

$$f(\vec{q},\vec{p}) = \frac{1}{(2\pi)^{2N}} \int M_{\mathrm{cm}}(X,\vec{\mu},\vec{\nu}) \exp\left[i(X - \vec{\mu}\vec{q} - \vec{\nu}\vec{p})\right] dX\, d\vec{\mu}\, d\vec{\nu}. \tag{57}$$

In quantum domain, the density operator is expressed in terms of the center-of-mass tomogram as follows:

$$\hat{\rho} = \frac{1}{(2\pi)^N} \int M_{\text{cm}}(X,\vec{\mu},\vec{v}) \exp\left[i\left(X - \vec{\mu}\hat{\vec{q}} - \vec{v}\hat{\vec{p}}\right)\right] dX\, d\vec{\mu}\, d\vec{v}. \qquad (58)$$

If one has symplectic tomogram of $N$-mode system, the symplectic tomogram of subsystem of $N-k$ modes reads

$$M(X_1, X_2, \ldots, X_{N-k}, \mu_1, \mu_2, \ldots, \mu_{N-k}, v_1, v_2, \ldots, v_{N-k})$$
$$= \int M(X_1, X_2, \ldots, X_N, \mu_1, \mu_2, \ldots, \mu_N, v_1, v_2, \ldots, v_N)\, dX_{N-k+1}\, dX_{N-k+2} \cdots dX_N. \qquad (59)$$

The center-of-mass tomogram of $N-k$ modes is expressed in terms of the center-of-mass tomogram of $N$ modes as

$$M_{\text{cm}}(X, \mu_1, \mu_2, \ldots, \mu_{N-k}, v_1, v_2, \ldots, v_{N-k}) = \int M_{\text{cm}}(X, \mu_1, \mu_2, \ldots, \mu_N, v_1, v_2, \ldots, v_N)$$
$$\times \delta(\mu_{N-k+1})\delta(\mu_{N-k+2})\cdots\delta(\mu_N)\, d\mu_{N-k+1}\, d\mu_{N-k+2} \cdots d\mu_N. \qquad (60)$$

The center-of-mass tomogram has the homogeneity property

$$M_{\text{cm}}(\lambda X, \lambda \vec{\mu}, \lambda \vec{v}) = \frac{1}{|\lambda|} M_{\text{cm}}(X, \vec{\mu}, \vec{v}). \qquad (61)$$

This means that

$$\left(\frac{\partial}{\partial X} + \vec{\mu}\frac{\partial}{\partial \vec{\mu}} + \vec{v}\frac{\partial}{\partial \vec{v}} + 1\right) M_{\text{cm}}(X, \vec{\mu}, \vec{v}) = 0. \qquad (62)$$

The center-of-mass tomogram, being nonnegative normalized homogeneous function determines a classical state if the integral in (57) is nonnegative. This tomogram determines the quantum state if the quantized integral in (58) is a nonnegative operator. These properties are analogous to the properties of symplectic tomograms.

There are center-of-mass tomograms which satisfy nonnegativity conditions for both integrals in (57) and (58). These tomograms correspond to Gaussians which satisfy the position–momentum uncertainty relations. Also there exist nonnegative, normalized, and homogeneous functions $M_{\text{cm}}(X, \vec{\mu}, \vec{v})$ which violate nonnegativity conditions for both (57) and (58). These tomograms determine neither classical nor quantum states.

## ENTROPIES IN THE PROBABILITY THEORY

In the information-theory context, entropy is related to an arbitrary probability-distribution function. We remind the notion of Shannon entropy [58].

Given the probability distribution $P(n)$, where $n$ is a discrete random variable, i.e., $P(n) \geq 0$, and the normalization condition holds $\sum_n P(n) = 1$, one has, by definition, the

Shannon entropy
$$S = -\sum_n P(n) \ln P(n) = -\langle \ln P(n) \rangle. \tag{63}$$

There exist other characteristics of the distribution $P(n)$, namely, an analog of the purity $\mu = \sum_n P^2(n)$ or the generalizations of this parameter $\mu(k) = \sum_n P^k(n)$.

There exist other kinds of entropies depending on extra parameter $q$ and associated to the probability distribution $P(n)$, for example, Rényi entropy [59]

$$R(q) = \frac{1}{1-q} \ln \left[ \sum_n \left( P(n) \right)^q \right], \qquad q > 0. \tag{64}$$

In the limit $q \to 1$, one has the equality of Rényi entropy to Shannon entropy

$$R(1) = S. \tag{65}$$

The Shannon entropy $S$ is a number. The Rényi entropy $R(q)$ is the function of a parameter $q$ and, due to this, it contains more information on details of the probability distribution $P(n)$ including the value of Shannon entropy given by (65).

For distribution functions $P(x)$, in the case of continuous random variables $x$, one has the same definition with the replacements $n \to x$ and $\sum_n \to \int dx$. For the case of quantum-system states with density matrix $\rho$, one has an analog of the Shannon entropy which is called von Neumann entropy given by formula with taking the trace

$$S_{vN} = -\text{Tr}(\rho \ln \rho) \tag{66}$$

and quantum Rényi entropy

$$R_\rho(q) = \frac{1}{1-q} \ln \text{Tr}(\rho^q). \tag{67}$$

The above formulas generalize the definition of entropies in the classical domain. In the limit $q \to 1$, for quantum system entropies, one has equality analogous to the previous equality (65), i.e., one has the equality of the quantum Réniy entropy to the von Neumann entropy

$$R_\rho(1) = S_{vN}. \tag{68}$$

Below we apply the introduced definitions of entropy in all cases of probability distributions and density matrices to tomograms, since the tomograms themselves are fair probability distributions for both classical and quantum systems.

## ENTROPIES OF CLASSICAL PARTICLE INCLUDING TOMOGRAPHIC ENTROPIES

In this section, we first apply the definition of Shannon entropy to the classical probability distributions on the phase space for systems with one and two degrees of freedom.

Given probability distributions $f_1(q_1, p_1)$ for one classical particle and $f_2(\vec{q}, \vec{p})$ for two classical particles on their phase spaces, where $\vec{q} = (q_1, q_2)$ and $\vec{p} = (p_1, p_2)$, one can introduce several marginal distributions

$$\mathcal{P}_1(q_1) = \int f_1(q_1, p_1) \, dp_1, \qquad \Pi_1(p_1) = \int f_1(q_1, p_1) \, dq_1,$$

$$\tilde{f}(q_1, p_1) = \int f_2(\vec{q}, \vec{p}) \, dq_2 \, dp_2, \qquad \tilde{f}(q_2, p_2) = \int f_2(\vec{q}, \vec{p}) \, dq_1 \, dp_1, \quad (69)$$

as well as

$$\tilde{\mathcal{P}}_1(q_1, q_2) = \int f_2(\vec{q}, \vec{p}) \, d\vec{p}, \qquad \tilde{\Pi}_1(p_1, p_2) = \int f_2(\vec{q}, \vec{p}) \, d\vec{q}. \quad (70)$$

The probability distributions determine the Shannon entropies

$$H_{12} = -\int f(\vec{q}, \vec{p}) \ln f(\vec{q}, \vec{p}) \, d\vec{q} \, d\vec{p},$$

$$H_{\mathcal{P}_1} = -\int \mathcal{P}_1(q_1) \ln \mathcal{P}_1(q_1) \, dq_1, \quad (71)$$

$$H_{\Pi_1} = -\int \Pi_1(p_1) \ln \Pi_1(p_1) \, dp_1.$$

We point out the following fact.

There is no special bounds for the sum $S = H_{\mathcal{P}_1} + H_{\Pi_1}$, which characterize the degree of order in the phase-space distribution described by the factorized probability density

$$f_{\mathcal{P}\Pi}(q_1, p_1) = \mathcal{P}_1(q_1) \Pi_1(p_1).$$

Analogously, if one considers two entropies for the systems with two degrees of freedom

$$\tilde{H}_{\tilde{\mathcal{P}}_1} = -\int \tilde{\mathcal{P}}_1(q_1, q_2) \ln \tilde{\mathcal{P}}_1(q_1, q_2) \, dq_1 \, dq_2,$$

$$\tilde{H}_{\tilde{\Pi}_1} = -\int \tilde{\Pi}_1(p_1, p_2) \ln \tilde{\Pi}_1(p_1, p_2) \, dp_1 \, qp_2, \quad (72)$$

there is no special bounds for the sum of the entropies $\tilde{S} = \tilde{H}_{\tilde{\mathcal{P}}_1} + \tilde{H}_{\tilde{\Pi}_1}$, which are the entropies associated with the factorized probability density

$$f_{\tilde{\mathcal{P}}_1 \tilde{\Pi}_1}(q_1, p_1, q_2, p_2) = \tilde{\mathcal{P}}_1(q_1, q_2) \tilde{\Pi}_1(p_1, p_2).$$

The absence of the entropic bounds for one and two degrees of freedom is connected with the absence of the corresponding bounds in the form of the uncertainty relations of quantum mechanics for the matrix elements of the nonnegative dispersion matrices associated with the probability distributions $f_1(q_1, p_1)$ and $f_2(\vec{q}, \vec{p})$.

The first dispersion matrix reads

$$\sigma = \begin{pmatrix} \sigma_{p_1 p_1} & \sigma_{p_1 q_1} \\ \sigma_{q_1 p_1} & \sigma_{q_1 q_1} \end{pmatrix}, \quad (73)$$

where one has variances of the momentum and position

$$\sigma_{p_1 p_1} = \int p_1^2 f_1(q_1, p_1) \, dq_1 \, dp_1 - \left( \int p_1 f_1(q_1, p_1) \, dq_1 \, dp_1 \right)^2,$$
$$\sigma_{q_1 q_1} = \int q_1^2 f_1(q_1, p_1) \, dq_1 \, dp_1 - \left( \int q_1 f_1(q_1, p_1) \, dq_1 \, dp_1 \right)^2, \quad (74)$$

and the covariance

$$\sigma_{q_1 p_1} = \sigma_{p_1 q_1}$$
$$= \int q_1 p_1 f_1(q_1, p_1) \, dq_1 \, dp_1 - \left( \int q_1 f_1(q_1, p_1) \, dq_1 \, dp_1 \right) \left( \int p_1 f_1(q_1, p_1) \, dq_1 \, dp_1 \right). \quad (75)$$

The nonnegativity of the dispersion matrix means that obvious inequalities hold, namely,

$$\sigma_{q_1 q_1} \geq 0, \qquad \sigma_{p_1 p_1} \geq 0, \qquad \sigma_{q_1 q_1} \sigma_{p_1 p_1} - \sigma_{q_1 p_1}^2 \geq 0. \quad (76)$$

Correspondingly, in the case of two degrees of freedom, for the second dispersion matrix

$$\widetilde{\sigma} = \begin{pmatrix} \sigma_{p_1 p_1} & \sigma_{p_1 q_1} & \sigma_{p_1 p_2} & \sigma_{p_1 q_2} \\ \sigma_{q_1 p_1} & \sigma_{q_1 q_1} & \sigma_{q_1 p_2} & \sigma_{q_1 q_2} \\ \sigma_{p_2 p_1} & \sigma_{p_2 q_1} & \sigma_{p_2 p_2} & \sigma_{p_2 q_2} \\ \sigma_{q_2 p_1} & \sigma_{q_2 q_1} & \sigma_{q_2 p_2} & \sigma_{q_2 q_2} \end{pmatrix}, \quad (77)$$

the nonnegativity condition provides, in addition to inequalities (76), the condition $\det \widetilde{\sigma} \geq 0$ and the property $\det \begin{pmatrix} \sigma_{p_1 p_1} & \sigma_{p_1 q_1} & \sigma_{p_1 p_2} \\ \sigma_{q_1 p_1} & \sigma_{q_1 q_1} & \sigma_{q_1 p_2} \\ \sigma_{p_2 p_1} & \sigma_{p_2 q_1} & \sigma_{p_2 p_2} \end{pmatrix} \geq 0$, where the matrix elements of matrix (77) are the variances and covariances of the positions and momenta associated with the probability distribution $f_2(\vec{q}, \vec{p})$. It is known that for the quantum particle there exist Robertson multimode uncertainty relations [60] making constrains on the properties of the dispersion matrix $\widetilde{\sigma}$, different from the discussed classical ones (76).

One can calculate also the Rényi entropies associated with the probability distributions $\mathscr{P}_1(q_1)$ and $\Pi_1(p_1)$; they read

$$R_{\mathscr{P}_1} = \frac{1}{1-q} \ln \int \mathscr{P}_1^q(q_1) \, dq_1, \qquad R_{\Pi_1} = \frac{1}{1-q} \ln \int \Pi_1^q(p_1) \, dp_1. \quad (78)$$

What is worth noting, that for the classical system the entropies $R_{\mathscr{P}_1}$ and $R_{\Pi_1}$ do not correlate; i.e., there is no special bound providing a constrain on the sum of these entropies.

In the limit $q \to 1$, the Rényi entropies associated with the probability distributions $\mathscr{P}_1(q_1)$ and $\Pi_1(p_1)$ go, respectively, $R_{\mathscr{P}_1} \to H_{\mathscr{P}_1}$ and $R_{\Pi_1} \to H_{\Pi_1}$, to the corresponding Shannon entropies associated with the same probability distributions $\mathscr{P}_1(q_1)$ and $\Pi_1(p_1)$.

Using classical distributions $f_1(q_1, p_1)$ and $f_2(\vec{q}, \vec{p})$, one can introduce symplectic tomograms of the classical-particle states for one degree of freedom (see Equation (4)) and for two degrees of freedom

$$M(X_1, \mu_1, \nu_1, X_2, \mu_2, \nu_2) = \int f_2(\vec{q}, \vec{p}) \delta(X_1 - \mu_1 q_1 - \nu_1 p_1) \delta(X_2 - \mu_2 q_2 - \nu_2 p_2) \, d\vec{q} \, d\vec{p}. \tag{79}$$

The tomograms are the probability-distribution functions; so they determine the Shannon entropies for one and two degrees of freedom, respectively,

$$H_M(\mu, \nu) = -\int M(X, \mu, \nu) \ln M(X, \mu, \nu) \, dX, \tag{80}$$

$$\widetilde{H}_M(\vec{\mu}, \vec{\nu}) = -\int M(X_1, \mu_1, \nu_1, X_2, \mu_2, \nu_2) \ln M(X_1, \mu_1, \nu_1, X_2, \mu_2, \nu_2) \, dX_1 \, dX_2, \tag{81}$$

where $\vec{\mu} = (\mu_1, \mu_2)$ and $\vec{\nu} = (\nu_1, \nu_2)$.

One has
$$H_M(1, 0) = H_{\mathscr{P}_1} \qquad \text{and} \qquad H_M(0, 1) = H_{\Pi_1}.$$

In the case of one degree of freedom, for $\mu = \cos\theta$ and $\nu = \sin\theta$, the entropy (80) depends on the "local oscillator" phase $\theta$, namely,

$$H_M(\cos\theta, \sin\theta) = H(\theta).$$

It is worth pointing out that there is no special bounds for classical tomographic probability distribution (4). We mean that there is no bound for the sum of entropies

$$\mathscr{H}(\theta) = H(\theta) + H(\theta + \pi/2),$$

and we show below that the bounds appear for quantum particles.

## ENTROPY AND ENTROPIC UNCERTAINTY RELATIONS

In quantum mechanics, for continuous variables the wave function $\psi(x)$ provides the probability-distribution density of position

$$P(x) = |\psi(x)|^2. \tag{82}$$

The corresponding entropy $S_x$ reads (see, for example, [60])

$$S_x = -\int |\psi(x)|^2 \ln |\psi(x)|^2 \, dx. \tag{83}$$

In the momentum representation, the wave function reads

$$\widetilde{\psi}(p) = \frac{1}{\sqrt{2\pi}} \int \psi(x) e^{-ipx} dx. \tag{84}$$

Corresponding entropy $S_p$ related to the momentum-probability density $|\widetilde{\psi}(p)|^2$ is

$$S_p = -\int |\widetilde{\psi}(p)|^2 \ln |\widetilde{\psi}(p)|^2 dp. \tag{85}$$

From mathematical point of view, there exists correlation of entropies $S_x$ and $S_p$, since function $\psi(x)$ determines Fourier component $\widetilde{\psi}(p)$. This means that entropies $S_x$ and $S_p$ have to obey some constrains. These constrains are well-known entropic uncertainty relations (some inequalities).

For one-mode system, the inequalities read (see, e.g., [60])

$$S_x + S_p \geq \ln(\pi e). \tag{86}$$

For multimode quantum systems, the entropy uncertainty relation reads

$$S_{\vec{x}} + S_{\vec{p}} \geq N \ln(\pi e), \tag{87}$$

where $N$ is the number of degrees of freedom of the system and

$$\begin{aligned} S_{\vec{x}} &= -\int |\psi(\vec{x})|^2 \ln |\psi(\vec{x})|^2 d\vec{x}, \\ S_{\vec{p}} &= -\int |\widetilde{\psi}(\vec{p})|^2 \ln |\widetilde{\psi}(\vec{p})|^2 d\vec{p}. \end{aligned} \tag{88}$$

The functions $\psi(\vec{x})$ and $\widetilde{\psi}(\vec{p})$ are connected by the Fourier transform

$$\widetilde{\psi}(\vec{p}) = (2\pi)^{-N/2} \int \psi(\vec{x}) e^{-i\vec{p}\vec{x}} d\vec{x}, \tag{89}$$

and this is a reason why one has inequality (87) for the entropies.

For the Gaussian wave function corresponding to factorized coherent state of several ($N$) modes,

$$S_{\vec{x}} + S_{\vec{p}} = N \ln(\pi e). \tag{90}$$

Thus, the coherent states saturate inequality (87).

The bounds (86) and (87) exist for the quantum particle, but as we discussed for corresponding classical entropies there is no such bounds.

Quantum bounds (86) and (87) are compatible with the Schrödinger–Robertson uncertainty relations for the position and momentum of the quantum particle

$$\sigma_{q_1 q_1} \sigma_{p_1 p_1} - \sigma_{q_1 p_1}^2 \geq 1/4, \tag{91}$$

which differs from $\sigma_{q_1 q_1} \geq 0$, $\sigma_{p_1 p_1} \geq 0$, $\sigma_{q_1 q_1} \sigma_{p_1 p_1} - \sigma_{q_1 p_1}^2 \geq 0$ by the presence of the bound $1/4$.

The bound (87) is compatible with multimode quantum Robertson uncertainty relations known for the variances and covariance of the position and momenta written in the form of dispersion matrix (77) but with extra term $\pm i/2$ (two-mode case)

$$\widetilde{\sigma} = \begin{pmatrix} \sigma_{p_1 p_1} & \sigma_{p_1 q_1} - i/2 & \sigma_{p_1 p_2} & \sigma_{p_1 q_2} \\ \sigma_{q_1 p_1} + i/2 & \sigma_{q_1 q_1} & \sigma_{q_1 p_2} & \sigma_{q_1 q_2} \\ \sigma_{p_2 p_1} & \sigma_{p_2 q_1} & \sigma_{p_2 p_2} & \sigma_{p_2 q_2} - i/2 \\ \sigma_{q_2 p_1} & \sigma_{q_2 q_1} & \sigma_{q_2 p_2} + i/2 & \sigma_{q_2 q_2} \end{pmatrix}$$

and expressed as the condition of nonnegativity of this matrix, i.e., the matrix has only nonnegative eigenvalues.

## TOMOGRAPHIC ENTROPIES

Since the symplectic tomogram is the standard probability distribution, one can introduce entropy associated with tomogram of quantum state [61] or with tomogram of analytic signal [62].

Thus one has entropy as the function of two real variables $\mu$ and $\nu$

$$S(\mu, \nu) = -\int M(X, \mu, \nu) \ln M(X, \mu, \nu) \, dX. \tag{92}$$

We call this entropy symplectic entropy.

In view of the homogeneity and normalization conditions for tomogram, one has the additivity property

$$S(\lambda \mu, \lambda \nu) = S(\mu, \nu) + \ln |\lambda|. \tag{93}$$

Also one has optical tomographic entropy associated with optical tomogram $w(X, \theta)$ as

$$S(\theta) = -\int w(X, \theta) \ln w(X, \theta) \, dX, \tag{94}$$

and this entropy depends on local oscillator phase in experiments with measuring photon homodyne quadrature.

The quantum optical tomogram of the pure state is determined by the wave function as

$$w(X, \theta) = \left| \int \psi(y) \exp\left[ \frac{i}{2} \left( \cot \theta \, (y^2 + X^2) - \frac{2X}{\sin \theta} y \right) \right] \frac{dy}{\sqrt{2\pi i \sin \theta}} \right|^2, \tag{95}$$

where we used symplectic tomogram in form (15) for $\mu = \cos \theta$ and $\nu = \sin \theta$ in which $\mu$ and $\nu$ were put.

On the other hand, this tomogram formally equals to

$$w(X, \theta) = |\psi(X, \theta)|^2, \tag{96}$$

where the wave function reads

$$\psi(X, \theta) = \frac{1}{\sqrt{2\pi i \sin \theta}} \int \exp\left[ \frac{i}{2} \left( \cot \theta \, (y^2 + X^2) - \frac{2X}{\sin \theta} y \right) \right] \psi(y) \, dy, \tag{97}$$

being the fractional Fourier transform of the wave function $\psi(y)$. This wave function corresponds to the wave function of a harmonic oscillator with $\hbar = m = \omega = 1$ taken at the "time" moment $\theta$ provided the wave function at the initial time moment $\theta = 0$ equals to $\psi(y)$.

In view of the expression of tomogram in terms of the wave function (96) and (97), one has the entropic uncertainty relation in the form

$$S(\theta) + S(\theta + \pi/2) \geq \ln \pi e. \tag{98}$$

Here $S(\theta)$ is the tomographic Shannon entropy associated with optical tomogram (95) which is measured by homodyne detector.

Thus in (98) the known entropic uncertainty relation (86) is extended to any value of the local oscillator phase $\theta$.

We illustrate the entropic inequality (98) by the example of the harmonic oscillator's ground state with the wave function

$$\psi_0(x) = \pi^{-1/4} e^{-x^2/2} \tag{99}$$

written in dimensionless variables. Using (95) and calculating the integral for the function $\psi(y)$ given by (99), we obtain the tomogram of the oscillator's ground state in the Gaussian form

$$w(X, \theta) = \pi^{-1/2} e^{-X^2}. \tag{100}$$

The ground-state tomogram does not depend on the angle $\theta$.

In view of this, the entropy for the Gaussian distribution reads

$$S(\theta) = S(\theta + \pi/2) = \frac{1}{2} \ln \pi e.$$

The sum of these two entropies saturates the previous inequality (98).

## RÉNYI ENTROPIC INEQUALITIES

In recent paper [63], the new uncertainty relation was obtained for quantum Rényi entropy related to the probability distributions for position and momentum of quantum state with density operator $\hat{\rho}$. The uncertainty relation reads

$$\frac{1}{1-\alpha} \ln \left( \int_{-\infty}^{\infty} dp \, [\rho(p,p)]^\alpha \right) + \frac{1}{1-\beta} \ln \left( \int_{-\infty}^{\infty} dx \, [\rho(x,x)]^\beta \right)$$
$$\geq -\frac{1}{2(1-\alpha)} \ln \frac{\alpha}{\pi} - \frac{1}{2(1-\beta)} \ln \frac{\beta}{\pi}, \tag{101}$$

where positive parameters $\alpha$ and $\beta$ satisfy the constrain

$$(1/\alpha) + (1/\beta) = 2. \tag{102}$$

Here $\rho(x,x)$ and $\rho(p,p)$ are diagonal elements of the density matrix in the position and momentum representation, respectively, and they determine the corresponding probability distributions. Rényi entropies $R_\alpha$ and $R_\beta$ related to the momentum and position distributions, respectively, are just two terms on the left-hand side of (101).

For $\alpha, \beta \longrightarrow 1$, these entropies become Shannon entropies $S_p$ and $S_x$.

We illustrate this inequality by the example of the harmonic oscillator's ground state. In this case, one has the Rényi entropies

$$R_\alpha = \frac{\ln \pi}{2} - \frac{1}{2}\frac{\ln \alpha}{1-\alpha}, \qquad R_\beta = \frac{\ln \pi}{2} - \frac{1}{2}\frac{\ln \beta}{1-\beta}$$

which, in the limit $\alpha \to 1$ and $\beta \to 1$, go to $(\ln e)/2$. Also the sum of the entropies reads

$$R_\alpha + R_\beta = \ln \pi - \frac{1}{2}\frac{\ln \alpha}{1-\alpha} - \frac{1}{2}\frac{\ln \beta}{1-\beta}\,.$$

In view of (102), this sum equals to the right-hand side of the previous inequality (101).

Thus, the harmonic oscillator's ground state saturates this entropic inequality for quantum Rényi entropy.

Using the same argument that we employed to obtain inequality (98) for Shannon entropies, we arrive at the condition for optical tomogram (see, [64])

$$(q-1)\ln\left\{\int_{-\infty}^\infty dX\,[w(X,\theta+\pi/2)]^{1/(1-q)}\right\} + (q+1)\ln\left\{\int_{-\infty}^\infty dX\,[w(X,\theta)]^{1/(1+q)}\right\}$$
$$\geq (1/2)\left\{(q-1)\ln[\pi(1-q)] + (q+1)\ln[\pi(1+q)]\right\}, \tag{103}$$

where the parameter $q$ is defined by $\alpha = (1-q)^{-1}$.

This inequality can be checked experimentally if one measures the optical tomogram.

## SPIN TOMOGRAMS (QUBIT AND QUDIT TOMOGRAMS)

Given an $N$-dimensional space of states of spin system. One can interpret this space either as the state space for one particle with spin $j = (N-1)/2$ (qudit) or in the case of product representation of number $N = n_1 n_2 \cdots n_M$ as the space of multipartite spin system (multipartite qudit system) with $j_1 = (n_1-1)/2$, $j_2 = (n_2-1)/2, \ldots, j_M = (n_M-1)/2$.

The $N \times N$ density matrix $\rho$ of the quantum state can be represented by the unitary tomogram of the spin state [65].

In the case of spin state with $j = (N-1)/2$, the tomogram is defined by the relation

$$w(m,u) = \langle m \mid u\rho u^\dagger \mid m \rangle, \tag{104}$$

where $\rho$ is the density matrix, $u$ is $N \times N$ unitary matrix and semi-integers $m = -j, -j+1, \ldots, j$ are values of spin projection on the $z$ axis. The tomogram can be considered as the probability vector $\vec{w}(u)$ on the unitary group with components labeled by spin projection $m$. If the density matrix $\rho$ has eigenvalues $\rho_m$ and eigenvectors $\vec{u}_m$, the vector $\vec{w}(u)$ is expressed as

$$\vec{w}(u) = |uu_0|^2 \vec{\rho}, \tag{105}$$

where $u_0$ is unitary matrix diagonalizing the matrix $\rho$. It is constructed from eigenvectors $\vec{u}_m$ as columns. By definition

$$|v|^2_{jk} = |v_{jk}|^2. \tag{106}$$

Tomogram (104) is nonnegative probability distribution function of random spin projection variable satisfying the normalization condition

$$\sum_{m=-j}^{j} w(m,u) = 1 \tag{107}$$

and the equality

$$\int w(m,u)\, du = 1, \tag{108}$$

where $du$ is Haar measure on unitary group with normalization

$$\int du = 1. \tag{109}$$

The important property of tomogram $w(m,u)$ is that its connection with density matrix $\rho$ reads

$$\rho \leftrightarrow w(m,u).$$

This means that the quantum state is given if the tomogram is known [18, 19].

## QUANTUM FOURIER TRANSFORM

The symmetric unitary $N \times N$ matrix $F$ with matrix elements $F_{jk} = \frac{1}{\sqrt{N}} \exp\left(\frac{2\pi i}{N} jk\right)$ ($j,k = 0, 1, \ldots, N-1$) can be used to provide invertible map of normalized complex vector $\vec{a}$ with components $a_k$ onto complex vectors $\vec{a}^{(f)}$ with components $a_k^{(f)}$

$$a_k^{(f)} = \sum_{j=0}^{N-1} F_{kj} a_j, \qquad a_k = \sum_{j=0}^{N-1} (F^\dagger)_{kj} a_j^{(f)}, \tag{110}$$

where the matrix $F$ with matrix elements $F_{kj}$ satisfies the equality

$$F^N = 1. \tag{111}$$

The map (110) is called the quantum Fourier transform.

If one uses the usual labels for spin projection $m = -j, -j+1, \ldots, j$, the operator $\hat{F}$ for quantum Fourier transform can be defined as

$$\hat{F}\,|\,m\rangle = \sum_{m'=-j}^{j} F_{m'm}\,|\,m'\rangle, \tag{112}$$

where the symmetric matrix
$$F_{m'm} = \langle m' | \hat{F} | m \rangle \quad (113)$$
has the form
$$F_{m'm} = \frac{1}{\sqrt{N}} \begin{pmatrix} 1 & 1 & 1 & \cdots & 1 \\ 1 & a & a^2 & \cdots & a^{N-1} \\ 1 & a^2 & a^4 & \cdots & a^{N-2} \\ \cdots & \cdots & \cdots & \cdots & \cdots \\ 1 & a^{N-1} & a^{N-2} & \cdots & a \end{pmatrix}, \quad a = \exp\left(\frac{2\pi i}{N}\right). \quad (114)$$

Thus the unitary operator of quantum Fourier transform reads
$$\hat{F} = \sum_{m=-j}^{j} \sum_{m'=-j}^{j} F_{m'm} | m' \rangle \langle m |. \quad (115)$$

In view of $F^N = 1$, one has $(\hat{F})^N = \hat{1}$, where $\hat{1}$ is the identity operator.

## SHANNON AND RÉNYI SPIN TOMOGRAPHIC ENTROPIES

Following standard definitions of probability theory, one can introduce Shannon [58] spin tomographic entropy [66, 67] and Rényi [59] tomographic entropy [67].

The Shannon tomographic entropy is the function on unitary group
$$H_u = -\sum_{m=-j}^{j} w(m, u) \ln w(m, u). \quad (116)$$

The Rényi tomographic entropy is also the function on the unitary group and it depends on extra parameter $q$
$$R_u = \frac{1}{1-q} \ln \left( \sum_{m=-j}^{j} (w(m,u))^q \right). \quad (117)$$

For two spin tomograms $w_1(m,u)$ and $w_2(m,u)$, we define the relative tomographic $q$-entropy
$$H_q(w_1(u)|w_2(u)) = -\sum_{m=-j}^{j} w_1(m,u) \ln_q \frac{w_2(m,u)}{w_1(m,u)} \quad (118)$$
with
$$\ln_q x = \frac{x^{1-q} - 1}{1-q}, \quad x > 0, \quad q > 0, \quad \ln_{q \to 1} x = \ln x. \quad (119)$$

The relative tomographic $q$-entropy is nonnegative function for any admissible deformation parameter $q$. For $q \to 1$, $R_u \to H_u$ and the relative $q$-entropy becomes relative

entropy associated to the two tomographic-probability distributions

$$H(w_1(u)|w_2(u)) = -\sum_{m=-j}^{j} w_1(m,u) \ln \frac{w_2(m,u)}{w_1(m,u)}. \quad (120)$$

The minimum over unitary group of the Rényi tomographic entropy $R_u = \frac{1}{1-q} \ln \left( \sum_{m=-j}^{j} (w(m,u))^q \right)$ is equal to quantum Rényi tomographic entropy [67]

$$\min R_u = \frac{1}{1-q} \ln \operatorname{Tr} \rho^q. \quad (121)$$

The minimum over unitary group of the Shannon tomographic entropy (116) is equal to the von Neumann entropy [66, 67], i.e.,

$$\min H_u = -\operatorname{Tr} \rho \ln \rho. \quad (122)$$

## SHANNON ENTROPIC INEQUALITIES IN MEASURING NONCOMMUTATIVE OBSERVABLES

In this section, we review known entropic inequalities considered, for example, in [68, 69, 70, 71, 72]. The inequalities appear in the problem of measuring two observables $\hat{A}$ and $\hat{B}$ in finite Hilbert space.

Let the spectral decompositions of Hermitian operators $\hat{A}$ and $\hat{B}$ read

$$\hat{A} = \sum_k A_k |a_k\rangle\langle a_k|, \quad \hat{B} = \sum_k B_k |b_k\rangle\langle b_k|, \quad k = 1,\ldots,N, \quad (123)$$

where $A_k$ and $B_k$ are eigenvalues of the observables and $|a_k\rangle$ and $|b_k\rangle$ are their orthonormal systems of eigenvectors.

For pure state $|\psi\rangle$, one has two probability distributions

$$p_k = |\langle a_k | \psi \rangle|^2, \quad q_k = |\langle b_k | \psi \rangle|^2. \quad (124)$$

The corresponding Shannon entropies connected with these two distributions read

$$H_p = -\sum_k p_k \ln p_k \quad (125)$$

and

$$H_q = -\sum_k q_k \ln q_k. \quad (126)$$

They satisfy the inequality found in [73]

$$H_p + H_q \geq -2 \ln \frac{1}{2}(1+c), \quad (127)$$

where the bound $c$ is determined by maximum values of scalar product modulus

$$c = \max_{j,k} |\langle a_j | b_k \rangle|. \tag{128}$$

In [74] the inequality was proposed to be improved

$$H_p + H_q \geq -2\ln c \tag{129}$$

and in [68] it was proved.

The problem of finding bounds for sums of the entropies can be formulated as the problem of entropic inequalities associated with probability distributions given by unitary matrices $u$ in the form $|u_{jk}|^2$, where one of the indices is fixed and the other one labels the event probability.

For example, one has two unitary matrices $u_A$ and $u_B$. These matrices can be considered as matrices diagonalizing two Hermitian matrices $A$ and $B$ corresponding to operators $\hat{A}$ and $\hat{B}$ in a given basis $|jm\rangle$. This means that spin tomographic symbols of operators $\hat{A}$ and $\hat{B}$ written in the vector form read

$$\vec{w}_A(u) = |uu_A|^2 \vec{A}, \qquad \vec{w}_B(u) = |uu_B|^2 \vec{B},$$

where $\vec{A} = (A_{-j}, A_{-j+1}, \ldots, A_j)$ and $\vec{B} = (B_{-j}, B_{-j+1}, \ldots, B_j)$. Any pure state $|\psi\rangle\langle\psi|$ has the matrix form in the same basis determined by unitary matrix $u_0$ which diagonalizes the density matrix of this operator given in the same basis. Thus, one has two matrices, $|u_A u_0|^2$ and $|u_B u_0|^2$. The first columns of these matrices provide two probability distributions with the corresponding Shannon entropies. The sum of these entropies has the bound, namely,

$$-\sum |u_A u_0|^2_{1k} \ln |u_A u_0|^2_{1k} - \sum |u_B u_0|^2_{1k} \ln |u_B u_0|^2_{1k} \geq \text{const.}, \tag{130}$$

where we consider $u_A$ and $u_B$ to be fixed and $u_0$ as an arbitrary unitary matrix.

For the case of observables $\hat{A}$ and $\hat{B}$ with eigenvectors providing equality [75, 70]

$$\max |\langle a_i | b_j \rangle| = \frac{1}{\sqrt{N}}, \tag{131}$$

inequality (129) reads [70]

$$H_p + H_q \geq \ln N. \tag{132}$$

Thus the dimensionality of Hilbert space $N$ appears in the entropic inequality.

The problem of such vector bases is related to the properties of finite spin-state spaces [76, 77]. It was widely discussed, for example, in [78, 79, 80, 81, 82].

The entropic inequalities for Shannon entropy can be also obtained in studying the problem of measuring several noncommutative observables with orthonormal sets of eigenvectors which satisfy the condition (131) (see [70, 71]).

# KNOWN INEQUALITIES FOR BIPARTITE AND TRIPARTITE SYSTEMS

The tomographic entropies satisfy some known inequalities found in [67].

For example, if the spin system is bipartite, i.e., one has spin $j_1$ and $j_2$, the basis in tensor product space reads

$$|m_1 m_2\rangle = |m_1\rangle |m_2\rangle. \tag{133}$$

In this case, the tomogram is the joint-probability distribution of two random spin projections $m_1 = -j_1, -j_1+1, \ldots, j_1$ and $m_2 = -j_2, -j_2+1, \ldots, j_2$ depending on $(2j_1+1)(2j_2+1) \times (2j_1+1)(2j_2+1)$ unitary matrix $u$. The tomogram reads

$$w(m_1, m_2, u) = \langle m_1 m_2 | u\rho(1,2)u^\dagger | m_1 m_2\rangle, \tag{134}$$

where $\rho(1,2)$ is the density matrix of bipartite system state with matrix elements

$$\rho(1,2)_{m_1 m_2, m'_1 m'_2} = \langle m_1 m_2 | \rho(1,2) | m'_1 m'_2\rangle. \tag{135}$$

For the tomogram, one can introduce the Shannon entropy $H_{12}(u)$ as

$$H_{12}(u) = -\sum_{m_1=-j_1}^{j_1} \sum_{m_2=-j_2}^{j_2} w(m_1, m_2, u) \ln w(m_1, m_2, u), \tag{136}$$

which satisfies the subadditivity condition for all elements of unitary group

$$H_{12}(u) \leq H_1(u) + H_2(u), \tag{137}$$

where $H_1(u)$ and $H_2(u)$ are Shannon entropies associated with subsystem tomograms

$$w_1(m_1, u) = \sum_{m_2=-j_2}^{j_2} w(m_1, m_2, u) \tag{138}$$

and

$$w_2(m_2, u) = \sum_{m_1=-j_1}^{j_1} w(m_1, m_2, u) \tag{139}$$

as follows:

$$H_k(u) = -\sum_{m_k=-j_k}^{j_k} w_k(m_k, u) \ln w_k(m_k, u), \qquad k=1,2. \tag{140}$$

From this inequality, in view of the relation between the von Neumann and tomographic entropies, follows the known inequality [67], which is subadditivity condition for corresponding von Neumann entropy for bipartite system

$$S_{12} \leq S_1 + S_2, \tag{141}$$

where

$$S_k = -\operatorname{Tr}\rho_k \ln\rho_k, \qquad k=1,2 \tag{142}$$

and
$$\rho_1 = -\mathrm{Tr}_2\rho(1,2), \qquad \rho_2 = -\mathrm{Tr}_1\rho(1,2). \tag{143}$$

For tripartite spin system with spins $j_1$, $j_2$, $j_3$ and density matrix $\rho(1,2,3)$, the spin tomogram reads

$$w(m_1,m_2,m_3,u) = \langle m_1 m_2 m_3 \mid u\rho(1,2,3)u^\dagger \mid m_1 m_2 m_3 \rangle. \tag{144}$$

One associates with this tomogram the Shannon entropy $H_{123}(u)$. This entropy satisfies inequality, which is the strong subadditivity condition on the unitary group. It reads [67]

$$H_{123}(u) + H_2(u) \leq H_{12}(u) + H_{23}(u), \tag{145}$$

where

$$H_{123}(u) = -\sum_{m_1=-j_1}^{j_1}\sum_{m_2=-j_2}^{j_2}\sum_{m_3=-j_3}^{j_3} w(m_1,m_2,m_3,u)\ln w(m_1,m_2,m_3,u) \tag{146}$$

and entropies $H_{12}(u)$, $H_{23}(u)$, and $H_2(u)$ are defined by means of projected tomograms

$$w_{12}(m_1,m_2,u) = \sum_{m_3=-j_3}^{j_3} w(m_1,m_2,m_3,u), \tag{147}$$

$$w_{23}(m_2,m_3,u) = \sum_{m_1=-j_1}^{j_1} w(m_1,m_2,m_3,u), \tag{148}$$

$$w_2(m_2,u) = \sum_{m_1=-j_1}^{j_1} w_{12}(m_1,m_2,u). \tag{149}$$

Our new inequality (145) does not contradict the known strong subadditivity condition for von Neumann entropy presented in [83, 84]

$$S_{123} + S_2 \leq S_{12} + S_{23}, \tag{150}$$

where

$$S_{123} = -\mathrm{Tr}\rho_{123}\ln\rho_{123}, \tag{151}$$

and other entropies are von Neumann entropies for reduced density matrices $\rho(1,2) = \mathrm{Tr}_3\rho(1,2,3)$ and $\rho(2,3) = \mathrm{Tr}_1\rho(1,2,3)$.

Inequalities

$$H_{12}(u) \leq H_1(u) + H_2(u) \quad \text{and} \quad H_{123}(u) + H_2(u) \leq H_{12}(u) + H_{23}(u)$$

are new inequalities for composite quantum finite-dimensional systems obtained in [67].

# NEW INEQUALITIES FOR RÉNYI TOMOGRAPHIC ENTROPIES

We continue the study of tomographic entropies along the line of our work [67] and derive new inequalities for spin tomographic entropies related to quantum Fourier transform.

For $N$-dimensional space, the analog of uncertainty relation for the Rényi entropies was proposed in the form [63]

$$\frac{1}{1-\alpha} \ln \left( \sum_{k=1}^{N} \widetilde{p}_k^{\alpha} \right) + \frac{1}{1-\beta} \ln \left( \sum_{l=1}^{N} p_l^{\beta} \right) \geq \ln N, \tag{152}$$

where

$$\widetilde{p}_k = |\widetilde{a}_k|^2, \qquad p_l = |a_l|^2, \qquad \frac{1}{\alpha} + \frac{1}{\beta} = 2, \tag{153}$$

and the complex numbers $\widetilde{a}_k$ and $a_l$ are connected by the quantum Fourier transform

$$\widetilde{a}_k = \frac{1}{\sqrt{N}} \sum_{l=1}^{N} \exp\left( i \frac{2\pi k l}{N} \right) a_l. \tag{154}$$

Below we use the inequalities to obtain new inequalities for Shannon and Rényi entropies associated with unitary spin tomograms.

The spin tomogram of a state with density operator $\rho$ can be considered as a column probability vector on unitary group with the components $w_m(u)$. Then we can introduce another $N$-vector with components $p_m(u) = \sqrt{w_m(u)}$. Applying inequality (152) to these vectors and using the notation

$$\left| \sum_{m'=-j'}^{j} F_{mm'} \sqrt{w(m',u)} \right| = \sqrt{w_F(m,u)}, \tag{155}$$

where $F_{mm'}$ is given by (114) and $w_F(m,u)$ is the probability distribution, we obtain the following inequality:

$$\frac{1}{1-\alpha} \ln \left( \sum_{m=-j}^{j} w(m,u)^{\alpha} \right) + \frac{1}{1-\beta} \ln \left( \sum_{m=-j}^{j} w_F(m,u)^{\beta} \right) \geq \ln N. \tag{156}$$

Also using for pure state $|\psi\rangle$ the definition of spin tomogram, we obtain another similar inequality

$$\frac{1}{1-\alpha} \ln \left( \sum_{m=-j}^{j} w(m,u)^{\alpha} \right) + \frac{1}{1-\beta} \ln \left( \sum_{m=-j}^{j} w(m,Fu)^{\beta} \right) \geq \ln N, \tag{157}$$

where $F$ is the quantum Fourier transform matrix.

Thus one has for Rényi tomographic entropy (117) the inequality for each unitary matrix

$$R_\alpha(u) + R_\beta(Fu) \geq \ln N. \tag{158}$$

Thus the unitary spin tomogram of the particle with spin $j$ for the state with $N \times N$ density matrix $\rho$, where $N = 2j+1$, must satisfy inequality (157).

In the limit $\alpha \to 1$, $\beta \to 1$, one gets inequalities for Shannon entropy of the spin state

$$H(u) + H(Fu) \geq \ln N. \tag{159}$$

Another inequality reads
$$H(u) + H_F(u) \geq \ln N, \tag{160}$$

where $H_F(u)$ is the Shannon entropy associated with the probability distribution $w_F(m, u)$.

For the minimum value of the Shannon entropy realized for unitary matrix $u_0$, one has the von Neumann entropy

$$H(u_0) = S_{\text{vN}}. \tag{161}$$

Inequality (159) written for $u_0$ as

$$H(u_0) + H(Fu_0) \geq \ln N \tag{162}$$

provides the inequality for the von Neumann entropy

$$S_{\text{vN}} + S(Fu_0) \geq \ln N, \tag{163}$$

where $S(Fu_0)$ is a new function which we defined.

It has the following physical meaning.

If the density operator $\hat{\rho}$ of the quantum state of spin is given in the form of spectral decomposition

$$\hat{\rho} = \sum_{q=-j}^{j} \lambda_q \, |q\rangle\langle q|, \tag{164}$$

one can identify the eigenstate $|q\rangle$ of the density operator $\hat{\rho}$ with "position" state.

In the Schwinger approach developed in [75, 81, 78, 79, 80, 82, 85], the states

$$|p\rangle = \hat{F} \, |q\rangle, \tag{165}$$

where $\hat{F}$ is the Fourier transform operator, are interpreted as "momentum" eigenstates.

The matrix elements $\langle p \, | \, \hat{F} \, | \, q \rangle = F_{pq}$ provide the matrix $F$ which coincides with the Fourier transform matrix.

Thus we have the interpretation of the new inequality in the same manner as it was done in the case of continuous variables.

The new entropy $S(Fu_0)$ in the inequality (163) is the Shannon entropy for "momentum" distribution, if we identify the standard von Neumann entropy with Shannon entropy for "position" distribution.

Now we discuss the group average Shannon and Rényi entropies introduced in [67].

Due to invariance of Haar measure, one can conclude that the group average Shannon tomographic entropy satisfies the inequality

$$\bar{H} = \int H(u) \, du \geq \frac{1}{2} \ln N. \tag{166}$$

Also for group average Rényi entropy (117), one has the inequality

$$\bar{R}_{\alpha\beta} = \int R_\alpha(u)\,du + \int R_\beta(u)\,du \geq \ln N, \qquad \frac{1}{\alpha} + \frac{1}{\beta} = 2. \tag{167}$$

Let us consider $N \times N$-unitary matrix $u_{jk}$. One has the inequality

$$-\sum_{j=1}^{N} \left(|u_{jk}|^2 \ln |u_{jk}|^2 + |(Fu)_{jk}|^2 \ln |(Fu)_{jk}|^2\right) \geq \ln N \tag{168}$$

or

$$-\sum_{j=1}^{N}\sum_{k=1}^{N} \left(|u_{jk}|^2 \ln |u_{jk}|^2 + |(Fu)_{jk}|^2 \ln |(Fu)_{jk}|^2\right) \geq N\ln N, \tag{169}$$

where $F_{jk}$ is the Fourier transform matrix.

Integrating inequality (168) over the unitary group with Haar measure normalized as (108) one has the inequality

$$-\int \left(\sum_{j=1}^{N} |u_{jk}|^2 \ln |u_{jk}|^2\right) du \geq \frac{1}{2}\ln N. \tag{170}$$

We demonstrated on the example of qudit that for tomograms of the spin states connected by quantum Fourier transforms one has constrains in the form of inequalities for Shannon tomographic entropies.

One can demonstrate analogous constrains for Rényi tomographic entropies too.

## CONCLUSIONS

To conclude, we point out the main aspects of the tomographic probability approach for describing systems states in classical and quantum domains.

As we have demonstrated, the dynamic symmetry of the systems plays a key role in the probability representation. For continuous variables, the dynamic symmetry is the symplectic group. The symplectic transform of photon quadrature components yields the notion of homodyne quadrature which is measured in experiments where the Wigner function and density operator are reconstructed from the optical tomograms of the photon quantum states.

In quantum information, qubits and qudits are closely related with the unitary group which plays in this case the role of dynamic symmetry. The unitary matrices provide a possibility to construct the bistochastic matrices. The modulus squared of matrix elements of the unitary matrix gives the probability distributions. The entropies associated with these distributions obey the inequalities, which turn out to be entropic uncertainty relations giving some quantum bounds for characteristics of spin states. There is the subgroup of unitary matrices group with the nonzero matrix elements only in each $k$th column $\exp(i\varphi) = u_{jk}$. The matrices $|u|_{jk}^2$ for this subgroup can be embedded into semigroup of bistochastic matrices. The subgroup of unitary matrices can be used to construct special basis in finite Helbert space.

The tomographic-probability representation is, in fact, a new formulation of quantum mechanics. In this formulation, the notion of state is the same as the notion of state in classical statistical mechanics. The quantum state is associated with fair probability distribution. This probability distribution is connected with traditional notion of quantum state (density matrix or wave function) by invertible map. For continuous variables like the position and momentum, this map is formulated in terms of integral Radon transform of the Wigner function. For discrete variables like qubits or spins, the map is given in terms of transforms expressed through matrix elements of unitary matrices, i.e., in terms of matrix elements of irreducible representations of the $SU(2)$ group. Existence of such invertible maps provides the possibility to present all the aspects of quantum mechanics like transitions between the quantum states, uncertainty relations, entanglement phenomenon using the formalism of classical probability theory including the notions of entropy, information, etc. Thus, the description of quantum and classical statistics and studying quantum–classical relations become more transparent in the tomographic-probability representation.

In spite that in both classical and quantum domains the states are described by the same notion – tomogram – the sets of classical and quantum tomograms are different. The tomographic-probability distributions satisfy the different extra conditions for classical and quantum states. These conditions provide different bounds for classical and quantum states.

The classical and quantum evolution equations in classical and quantum domains yield the state orbits situated in the corresponding either classical or quantum sets (see related recent paper [86]). It is worth noting that the set of all tomographic-probability distributions contains not only classical and quantum states but also the tomograms which correspond neither classical nor quantum states. The use of these tomograms is not clear and needs to be investigated in detail. There exist some other realizations of the tomographic approach [87] based on generalized Radon transform which can be used in the formalism of classical and quantum statistics.

## Acknowledgments

This study was partially supported by the Russian Foundation for Basic Research under Projects Nos. 09-02-142 and 10-02-00312. The authors are grateful to the Organizers of the XL Latin American School of Physics *Symmetries in Physics* in memory of Marcos Moshinsky and especially to Prof. Octavio Castaños for kind hospitality.